\documentclass[runningheads]{llncs}
\usepackage[T1]{fontenc}
\usepackage{pifont} 
\usepackage{tikz}
\usetikzlibrary{shapes,arrows,positioning,fit,backgrounds,calc}
\usepackage{tikz}
\usetikzlibrary{arrows.meta, positioning, shapes.geometric, calc}

\usepackage{graphicx,verbatim}
\usepackage{amsmath}
\usepackage{amssymb}
\begin{document}
\title{Clinically-aligned ischemic stroke segmentation and ASPECTS scoring on NCCT imaging using a slice-gated loss on foundation representations}
%

\author{Hiba Azeem, Behraj Khan , Tahir Qasim Syed}  
\authorrunning{Anonymized Author et al.}
\institute{Institute of Business Administration Karachi \\
    \email{tahirqsyed@gmail.com}}
  
\maketitle              
\begin{abstract}
Rapid infarct assessment on non-contrast CT (NCCT) is essential for acute ischemic stroke management. Most deep learning methods perform pixel-wise segmentation without modeling the structured anatomical reasoning underlying ASPECTS scoring, where basal ganglia (BG) and supraganglionic (SG) levels are clinically interpreted in a coupled manner. We propose a clinically aligned framework that combines a frozen DINOv3 backbone with a lightweight decoder and introduce a Territory-Aware Gated Loss (TAGL) to enforce BG–SG consistency during training. This anatomically informed supervision adds no inference-time complexity. Our method achieves a Dice score of 0.6385 on AISD, outperforming prior CNN and foundation-model baselines. On a proprietary ASPECTS dataset, TAGL improves mean Dice from 0.698 to 0.767. These results demonstrate that integrating foundation representations with structured clinical priors improves NCCT stroke segmentation and ASPECTS delineation.
\end{abstract}
\section{Introduction}

Acute ischemic stroke (AIS) is a leading cause of mortality and long-term disability worldwide \cite{feigin2025world}. Clinical outcome critically depends on rapid diagnosis and timely intervention, including intravenous thrombolysis and endovascular thrombectomy \cite{atlantis2004association,ciccone2013endovascular}. In emergency settings, non-contrast computed tomography (NCCT) is the first-line imaging modality due to its speed, availability, and ability to exclude hemorrhage \cite{mainali2014detection}. However, early ischemic changes on NCCT are subtle and low-contrast, making reliable infarct delineation challenging even for experienced radiologists.

\noindent Recent deep learning methods have shown promising results for NCCT infarct segmentation, yet several challenges remain. Many high-performing approaches rely on hybrid CNN--Transformer architectures or fully fine-tuned large models, incurring substantial computational overhead that limits applicability in time-critical workflows \cite{hatamizadeh2022unetr,kuang2024hybrid}. Moreover, the scarcity of large-scale pixel-level annotated NCCT datasets constrains supervised learning and limits generalization, particularly for data-hungry architectures trained end-to-end \cite{liang2021symmetry}. Finally, most existing methods treat infarct segmentation as an independent pixel-wise task, without explicitly modeling the structured anatomical reasoning used in clinical stroke assessment.

\noindent In clinical practice, radiologists assess ischemic damage using the Alberta Stroke Program Early CT Score (ASPECTS), which partitions the middle cerebral artery territory into anatomically defined regions across two axial levels: basal ganglia (BG) and supraganglionic (SG). Importantly, these regions are not interpreted independently. Findings in the BG slice often guide interpretation at the SG level, reflecting anatomical proximity and vascular continuity. Despite extensive work on automated ASPECTS scoring and infarct segmentation \cite{kuang2019automated,kuang2021eis,wei2025deep}, this clinically grounded BG--SG coupling is largely absent from existing learning-based frameworks.

\noindent Foundation models pretrained via large-scale self-supervised learning provide an attractive alternative to task-specific architectures. Vision Transformers trained with DINO-style objectives yield transferable representations that can be reused for dense prediction tasks with minimal task-specific supervision \cite{bommasani2021opportunities,simeoni2025dinov3}. Recent studies demonstrate that frozen DINO backbones paired with lightweight decoders achieve strong NCCT segmentation performance without costly end-to-end fine-tuning \cite{yang2025segdino,zhang2025benchmarking}. However, incorporating clinically meaningful anatomical constraints into such foundation-model-based pipelines remains an open challenge.

\noindent In this work, we propose an efficient and anatomically informed framework for NCCT infarct segmentation and ASPECTS region delineation. Our approach employs a frozen DINOv3 backbone with a lightweight DPT-style decoder \cite{ranftl2021vision}. To explicitly encode clinical reasoning, we introduce a \emph{Territory-Aware Gated Loss} (TAGL) that enforces consistency between BG and SG predictions by selectively propagating information from the BG slice when infarction evidence is present. This introduces a structured inductive bias aligned with radiologist practice, without increasing inference-time complexity.

\noindent We evaluate our method on the public AISD dataset \cite{liang2021symmetry} and a proprietary NCCT dataset with expert-annotated ASPECTS regions. Results demonstrate strong segmentation performance with high computational efficiency, and consistent improvements in ASPECTS delineation when BG--SG coupling is enforced.
\section{Related Work}

\noindent \textbf{NCCT Infarct Segmentation.}
Early approaches to NCCT infarct segmentation relied on convolutional encoder--decoder architectures, with U-Net variants serving as a common baseline \cite{ronneberger2015u}. Subsequent work addressed low contrast and class imbalance through symmetry modeling, attention mechanisms, and multi-scale feature aggregation \cite{liang2021symmetry,kuang2021eis,ni2022asymmetry}. While effective, these CNN-based methods remain limited by local receptive fields, restricting long-range contextual reasoning required for anatomically structured infarcts.

\noindent \textbf{CNN--Transformer Hybrids.}
To capture global context, several methods introduced hybrid CNN--Transformer architectures, such as UNETR and CoTr \cite{hatamizadeh2022unetr,xie2021cotr}. In NCCT stroke imaging, designs incorporating bilateral or circular feature modeling further improved performance \cite{kuang2024hybrid}. However, these models typically incur increased computational cost and parameter count, limiting their practicality in time-sensitive clinical settings.

\noindent \textbf{ASPECTS Automation.}
Automated ASPECTS scoring has evolved from classical machine learning pipelines to deep learning-based segmentation and scoring frameworks. Early methods relied on handcrafted features and region-wise classifiers \cite{kuang2019automated}, while later approaches integrated voxel-wise segmentation with region aggregation in multi-task CNN architectures \cite{kuang2021eis}. Recent systems adopt staged pipelines combining localization, segmentation, and scoring \cite{wei2025deep}. Despite these advances, most approaches treat ASPECTS regions independently or rely on atlas-based aggregation, without explicitly encoding BG--SG anatomical coupling.

\noindent \textbf{Foundation Models for Medical Segmentation.}
Self-supervised foundation models have recently demonstrated strong transferability to medical image segmentation. Vision Transformers pretrained with DINO-style objectives produce general-purpose representations that perform well under limited supervision \cite{bommasani2021opportunities,simeoni2025dinov3}. Frozen DINO backbones combined with lightweight decoders achieve competitive NCCT segmentation performance without extensive fine-tuning \cite{yang2025segdino,zhang2025benchmarking}. While adapter-based or partially unfrozen variants further improve accuracy, they reintroduce additional parameters and training complexity \cite{chen2026strdiseg}.

\noindent \textbf{Our Positioning.}
In contrast to prior work, we combine the efficiency of frozen foundation models with clinically informed anatomical reasoning. Unlike CNN-based and hybrid architectures, we avoid heavy encoders and end-to-end fine-tuning. Unlike existing ASPECTS automation frameworks, we explicitly encode BG--SG coupling through a territory-aware loss rather than post hoc aggregation or atlas constraints. This enables principled integration of foundation-model representations with domain-specific inductive bias for NCCT stroke analysis.

\section{Method}
\begin{figure}[t]
\centering
\resizebox{\textwidth}{!}{
\begin{tikzpicture}[
    img/.style = {rectangle, draw, thick, text width=2.7cm, align=center,
                  minimum height=2.3cm, fill=white, font=\small},
    backbone/.style = {rectangle, draw, rounded corners,
                      text width=2.6cm, align=center,
                      minimum height=1.0cm, fill=green!10, font=\small},
    decoder/.style = {rectangle, draw, rounded corners,
                     text width=3.0cm, align=center,
                     minimum height=1.0cm, fill=orange!10, font=\small},
    aux/.style = {rectangle, draw, rounded corners,
                  text width=2.8cm, align=center,
                  minimum height=0.85cm, fill=green!10, font=\small},
    loss/.style = {rectangle, draw, dashed, rounded corners,
                  text width=3.2cm, align=center,
                  minimum height=1.0cm, fill=red!5, font=\small},
    pred/.style = {rectangle, draw, thick,
                  text width=2.8cm, align=center,
                  minimum height=0.95cm, fill=gray!10, font=\small},
    arrow/.style      = {thick,      ->, >=stealth},
    thickarrow/.style = {very thick, ->, >=stealth},
]


\node[img]      (input)  at (0,    0)   {\includegraphics[width=1.8cm]{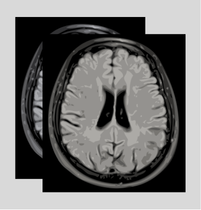}\\SG Slice $X$};
\node[backbone] (dino)   at (4,    0)   {DINOv3};

\node[img]      (input2) at (0,    3.8) {\includegraphics[width=1.8cm]{brain.png}\\BG Slice $X$};
\node[aux]      (msf)    at (4,    3.8) {DINOv3};

\node[decoder]  (dpt1)   at (8.5,  3.8) {DPT-Style\\Dual-Arm Decoder};
\node[decoder]  (dpt2)   at (8.5,  0)   {DPT-Style\\Dual-Arm Decoder};

\node[loss]     (tagl)   at (12.5, 1.9) {Territory-Aware\\Gated Loss\\(ASPECTS only)};

\node[pred]     (predn)  at (16.5, 1.9) {Segmentation Head};

\draw[arrow] (input.east)  -- (dino.west);

\draw[arrow] (input2.east) -- (msf.west);

\draw[arrow] (msf.east) -- (dpt1.west);

\draw[arrow] (dino.east) -- (dpt2.west);

\draw[arrow]
(dpt1.east) -- ++(0.8,0)                
            |- (tagl.north);            

\draw[arrow]
(dpt2.east) -- ++(0.8,0)                
            |- (tagl.south);            

\draw[arrow] (tagl.east) -- (predn.west);

\end{tikzpicture}
}
\caption{Overview of the proposed framework. An NCCT slice is processed by a
frozen DINOv3 backbone and a multi-scale feature extractor in parallel.
Two DPT-style dual-arm decoders process the respective features, and their
outputs are fused via a Territory-Aware Gated Loss (TAGL) module before
producing the final multi-class segmentation map $\hat{Y}$.}
\label{fig:method_overview_horizontal}
\end{figure}

Fig.~\ref{fig:method_overview_horizontal} illustrates the proposed framework. Our design follows two principles: (i) leverage strong pretrained representations from a foundation model without extensive architectural modification, and (ii) inject clinically motivated inductive bias only at the supervision level.

\noindent\textbf{Problem Formulation.}
Given an NCCT slice $X \in \mathbb{R}^{H \times W}$, the goal is to predict a pixel-wise label map $Y \in \{0,\dots,C-1\}^{H \times W}$, where $C$ denotes the number of classes. We consider three settings: binary infarct segmentation ($C=2$), multi-class AISD segmentation ($C=K$ anatomical regions), and ASPECTS segmentation ($C=11$, background plus ten territories). The model learns a mapping $f_\theta: X \rightarrow \hat{Y}$, producing class-wise logits $\hat{Y} \in \mathbb{R}^{C \times H \times W}$, with class probabilities obtained via softmax.

\noindent\textbf{Model Overview.}
We adopt a frozen Vision Transformer pretrained via self-supervised learning as the feature extractor and pair it with a lightweight dense prediction decoder to produce high-resolution segmentation maps. The backbone remains fixed during training, while only the decoder parameters are optimized. This design preserves the generalization capability of pretrained representations while keeping training and inference efficient. Further architectural details are provided in Fig.~\ref{fig:method_overview_horizontal}.

\noindent \textbf{Territory-Aware Gated Loss.}
Standard segmentation losses treat pixels independently and do not encode anatomical dependencies across axial levels. However, in ASPECTS scoring, evidence of infarction in the basal ganglia (BG) often implies involvement in corresponding supraganglionic (SG) regions. To encode this clinical prior, we introduce a Territory-Aware Gated Loss (TAGL). Let $\mathbf{p}^{\text{BG}}, \mathbf{p}^{\text{SG}} \in [0,1]^{H \times W}$ denote the predicted infarct probability maps for aligned BG and SG slices.

\noindent\textbf{Gate.}
The loss is activated only at locations where BG predictions exceed a small threshold $\tau$, using a binary gate
$g_{ij} = \mathbb{1}(p^{\text{BG}}_{ij} > \tau)$.

\noindent\textbf{Adaptive Weight.}
To modulate the penalty strength, we compute slice-level confidence scores
$C_{\text{BG}} = \frac{1}{HW}\sum_{i,j} p^{\text{BG}}_{ij}$ and
$C_{\text{SG}} = \frac{1}{HW}\sum_{i,j} p^{\text{SG}}_{ij}$.
An adaptive weight is then obtained via softmax,
$q = \frac{\exp(C_{\text{BG}})}{\exp(C_{\text{BG}}) + \exp(C_{\text{SG}})}$,
which emphasizes BG-driven supervision when BG confidence is higher.

\noindent\textbf{Disagreement Penalty.}
We penalize inconsistent predictions across aligned locations using the absolute difference
$d_{ij} = |p^{\text{SG}}_{ij} - p^{\text{BG}}_{ij}|$.

\noindent\textbf{Loss Definition.}
The per-pixel territory-aware penalty is defined as
$\mathcal{L}^{\text{TA}}_{ij} = g_{ij} \cdot q \cdot d_{ij}$,
and the overall loss is obtained by spatial averaging,
\begin{equation} 
\mathcal{L}^{\text{TA}} = \frac{1}{HW}\sum_{i,j} \mathcal{L}^{\text{TA}}_{ij}
\end{equation}
\noindent\textbf{Training Objective.}
For ASPECTS, the final objective is
\begin{equation} 
\mathcal{L}_{\text{total}} = \mathcal{L}_{\text{seg}} + \lambda \mathcal{L}^{\text{TA}}
\end{equation}
where $\mathcal{L}_{\text{seg}}$ is a standard segmentation loss and $\lambda$ balances the two terms. For AISD, where paired BG--SG supervision is unavailable, only $\mathcal{L}_{\text{seg}}$ is used.

\section{Experiments}

We evaluate the proposed framework on multiple NCCT infarct segmentation benchmarks to assess both general segmentation performance and the effect of clinically informed anatomical coupling. Our experiments are designed to answer three questions: (i) how effective frozen foundation models are for NCCT segmentation, (ii) whether a lightweight decoder suffices when paired with strong pretrained features, and (iii) whether territory-aware supervision improves ASPECTS-consistent predictions.

\textbf{Datasets}
We conduct experiments on two datasets with complementary characteristics.

\paragraph{AISD.}
AISD is a publicly available NCCT dataset with multi-class region-level annotations. It provides anatomically labeled infarct regions but does not include paired basal ganglia (BG) and supraganglionic (SG) slice supervision. AISD is used to evaluate general multi-class segmentation performance without anatomical coupling.

\paragraph{Proprietary ASPECTS Dataset.}
We additionally evaluate on an in-house ASPECTS dataset annotated according to the ten ASPECTS territories. Each case includes paired BG and SG slices, enabling supervision that reflects clinical scoring practice. This dataset is used to assess the effectiveness of the proposed Territory-Aware Gated Loss.

\textbf{Implementation Details}
All experiments are conducted using PyTorch. Input NCCT slices are resized to a fixed spatial resolution and intensity-normalized following standard clinical preprocessing. Data augmentation includes random horizontal flipping and mild intensity perturbations.

We use DINOv3-small as a frozen backbone throughout all experiments. Features are extracted from a fixed set of transformer layers and passed to a lightweight DPT-style decoder. Only the decoder parameters are optimized during training.

Models are trained using the AdamW optimizer with a fixed learning rate. Training is performed for a fixed number of epochs, and the model with the best validation performance is selected for evaluation. Unless otherwise stated, all hyperparameters are kept identical across datasets to ensure fair comparison.

\textbf{Training Objectives}
For AISD, we optimize a standard segmentation loss $\mathcal{L}_{\text{seg}}$, implemented as a combination of cross-entropy and Dice loss.

For the ASPECTS dataset, the training objective additionally includes the proposed Territory-Aware Gated Loss:
\begin{equation}
\mathcal{L}_{\text{total}} = \mathcal{L}_{\text{seg}} + \lambda \mathcal{L}^{\text{TA}},
\end{equation}
where $\lambda$ controls the contribution of anatomical coupling. The threshold $\tau$ used in the gating mechanism is fixed across all experiments.

\textbf{Evaluation Metrics}
We report Dice coefficient and mean Intersection-over-Union (mIoU) for segmentation quality. For ASPECTS, we additionally report region-wise Dice scores to assess performance across individual territories. All results are averaged over the test set.

\textbf{Baselines}
We compare our approach against representative CNN-based segmentation models and recent foundation-model-based methods. For fair comparison, all baselines are evaluated using their officially reported settings or re-implemented under identical preprocessing and evaluation protocols when necessary.

\textbf{Ablation Studies}
To isolate the contribution of each component, we conduct ablation experiments on the ASPECTS dataset by: (i) removing the territory-aware loss, (ii) disabling the adaptive confidence weighting, and (iii) replacing the frozen backbone with a partially fine-tuned variant. These studies quantify the effect of anatomical coupling and backbone freezing on segmentation performance.

\section{Results}

We report quantitative and qualitative results on AISD and the proprietary ASPECTS dataset, evaluating both segmentation accuracy and the impact of anatomically informed supervision.

\textbf{Results on AISD}
Table~\ref{tab:aisd_comparison} summarizes performance on the AISD binary infarct segmentation task. Using a hybrid BCE--Dice loss yields a substantial improvement over BCE alone, confirming the importance of overlap-based objectives under severe class imbalance. Our final model achieves a Dice score of \textbf{0.6385}, representing a large margin over previously reported results.

Compared to the frozen DINOv3 benchmarking baseline (Dice $=0.3674$), our approach nearly doubles segmentation accuracy while retaining a lightweight architecture. It also outperforms StrDiSeg with an unfrozen backbone (Dice $=0.5160$), demonstrating that effective use of pretrained representations does not require extensive fine-tuning when paired with an appropriate decoder and training objective.

\begin{figure}[!ht]
\centering
\includegraphics[width=0.5\textwidth]{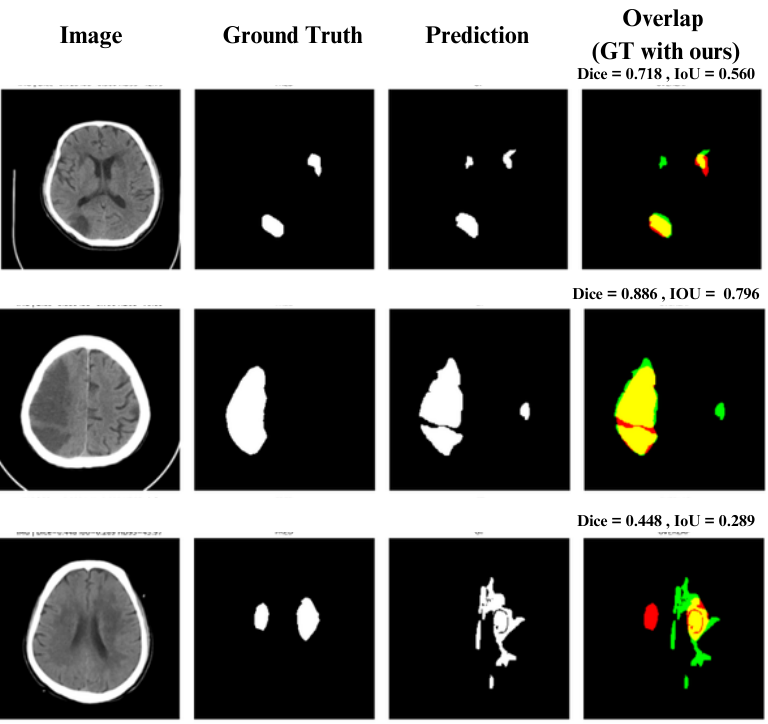}
\caption{Qualitative segmentation results on the AISD dataset produced by our best-performing model. 
Examples illustrate typical lesion appearances across challenging NCCT slices, including low-contrast regions and small or fragmented infarcts.}
\label{fig:aisd_visuals}
\end{figure}

Qualitative examples are shown in Fig.~\ref{fig:aisd_visuals}. The model consistently delineates ischemic regions across challenging NCCT slices, including low-contrast lesions and fragmented infarcts, indicating robustness to appearance variability.

\begin{figure}[!ht]
\centering
\includegraphics[width=0.48\textwidth]{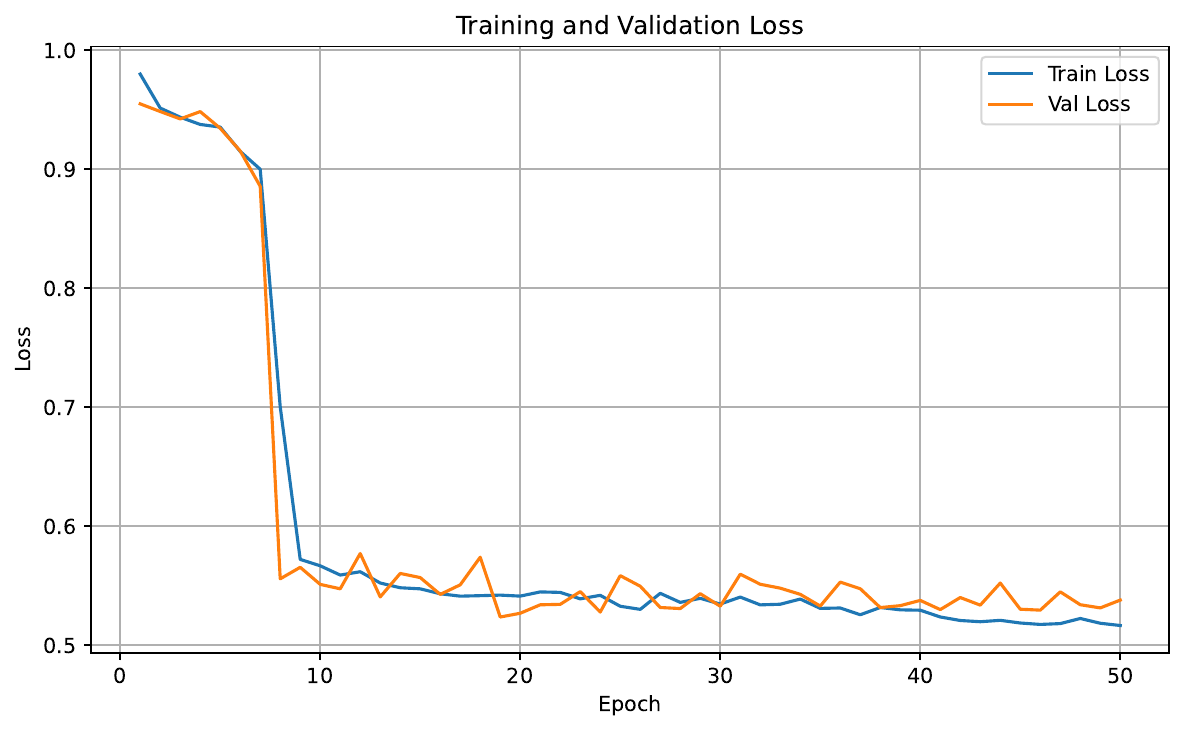}
\includegraphics[width=0.48\textwidth]{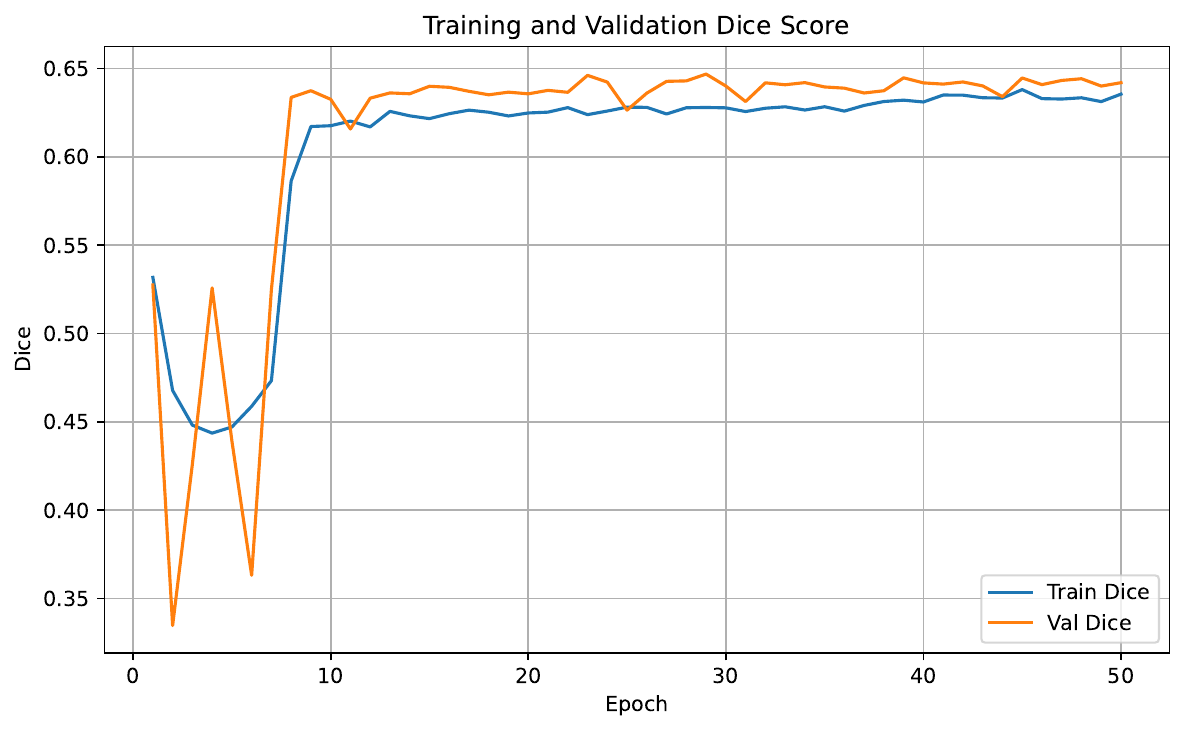}
\caption{Training and validation performance curves. 
Left: Combined BCE–Dice loss. 
Right: Corresponding Dice score evolution. 
Model selection is based on the highest validation Dice.}
\label{fig:training_curves}
\end{figure}

\textbf{Training Dynamics}
Fig.~\ref{fig:training_curves} illustrates training and validation behavior on AISD. Both loss and Dice curves exhibit rapid convergence followed by stable saturation, with a minimal gap between training and validation performance. This suggests effective optimization and good generalization. Improvements beyond approximately 30 epochs are marginal, indicating convergence under the adopted training configuration.

\begin{table}[!ht]
\caption{Performance comparison on AISD. Best results in bold.}
\label{tab:aisd_comparison}
\centering
\begin{tabular}{l l c c}
\hline
\textbf{Model} & \textbf{Encoder Type} & \textbf{Dice} & \textbf{IoU} \\
\hline
SEAN \cite{liang2021symmetry} & CNN & 0.5784 & -- \\
UNETR \cite{hatamizadeh2022unetr} & Swin Transformer & 0.4470 & 0.3510 \\
SwinUNETR \cite{hatamizadeh2021swin} & Swin Transformer & 0.4470 & 0.3510 \\
UNet++ \cite{zhou2019unet++} & CNN & 0.4410 & 0.3520 \\
DINOv3 benchmarking \cite{zhang2025benchmarking} & Frozen DINOv3 & 0.3674 & 0.2703 \\
StrDiSeg (frozen) \cite{chen2026strdiseg} & Frozen DINOv3 & 0.3720 & 0.2690 \\
StrDiSeg (unfrozen) \cite{chen2026strdiseg} & Unfrozen DINOv3 & 0.5160 & 0.4000 \\
\textbf{Ours} & Unfrozen DINOv3 & \textbf{0.6385} & \textbf{0.6231} \\
\hline
\end{tabular}
\end{table}

\textbf{Results on ASPECTS}
We next evaluate the effect of anatomical coupling on the proprietary ASPECTS dataset. Results are reported in Table~\ref{tab:aspects_comparison}, where metrics are averaged over all 11 classes (background plus ten ASPECTS territories).

The baseline model trained with standard cross-entropy loss achieves a mean Dice score of 0.7668. Incorporating the proposed Territory-Aware Gated Loss leads to a consistent improvement in both Dice and IoU, increasing mean Dice to \textbf{0.767}. This gain confirms that explicitly enforcing consistency between basal ganglia and supraganglionic predictions provides a useful inductive bias for ASPECTS segmentation beyond pixel-wise supervision alone.

\begin{table}[!ht]
\caption{Performance comparison on the proprietary ASPECTS dataset. Best results in bold.}
\label{tab:aspects_comparison}
\centering
\begin{tabular}{l c c}
\hline
\textbf{Configuration} & \textbf{Mean Dice} & \textbf{IoU} \\
\hline
Baseline (CE only)      & 0.698 & 0.683 \\
Proposed (CE + TAGL)    & \textbf{0.767} & \textbf{0.756} \\
\hline
\end{tabular}
\end{table}
\section{Conclusion}

We presented a clinically aligned framework for ischemic stroke segmentation and ASPECTS delineation on NCCT imaging that combines frozen foundation representations with anatomically informed supervision. By leveraging DINOv3 features and a lightweight decoder, our approach achieves strong segmentation accuracy while maintaining computational efficiency and avoiding extensive backbone fine-tuning.

To bridge the gap between pixel-wise learning and clinical reasoning, we introduced a Territory-Aware Gated Loss that explicitly models the coupling between basal ganglia and supraganglionic territories. This structured inductive bias improves ASPECTS-consistent predictions without altering the inference pipeline. Experimental results demonstrate state-of-the-art performance on AISD and significant gains on a proprietary ASPECTS dataset when anatomical coupling is enforced.

Overall, our findings suggest that integrating foundation models with clinically grounded constraints offers a principled and effective direction for robust, interpretable, and deployment-ready NCCT stroke analysis systems.

\bibliographystyle{splncs04}
\bibliography{arxiv}



\end{document}